\documentclass[conference]{IEEEtran}
\usepackage{graphicx} 
\usepackage{amsmath}
\usepackage{comment}
\usepackage{adjustbox}
\usepackage{graphicx}
\usepackage{booktabs}
\usepackage{svg}
\usepackage{subcaption}
\usepackage{threeparttable}
\usepackage{multirow}
\usepackage{url}
\usepackage{soul}

\usepackage{tikz}

\newcommand\copyrighttext{%
  \footnotesize \textcopyright \the\year{} IEEE. Personal use of this material is permitted. Permission from IEEE must be obtained for all other uses, including reprinting/republishing this material for advertising or promotional purposes, collecting new collected works for resale or redistribution to servers or lists, or reuse of any copyrighted component of this work in other works.}

\newcommand\copyrightnotice{%
\begin{tikzpicture}[remember picture,overlay]
\node[anchor=south,yshift=10pt] at (current page.south) {\fbox{\parbox{\dimexpr0.75\textwidth-\fboxsep-\fboxrule\relax}{\copyrighttext}}};
\end{tikzpicture}%
}

\title{Multi-Label Requirements Classification with Large Taxonomies}

\author{\IEEEauthorblockN{1\textsuperscript{st} Waleed Abdeen}
\IEEEauthorblockA{\textit{Blekinge Institute of Technology}\\
Karlskrona, Sweden \\
waleed.abdeen@bth.se}
\and
\IEEEauthorblockN{2\textsuperscript{nd} Michael Unterkalmsteiner}
\IEEEauthorblockA{\textit{Blekinge Institute of Technology}\\
Karlskrona, Sweden \\
michael.unterkalmsteiner@bth.se}
\and
\IEEEauthorblockN{3\textsuperscript{th} Krzysztof Wnuk}
\IEEEauthorblockA{\textit{Blekinge Institute of Technology}\\
Karlskrona, Sweden \\
krzysztof.wnuk@bth.se}
\and
\IEEEauthorblockN{4\textsuperscript{rd} Alexandros Chirtoglou}
\IEEEauthorblockA{\textit{HOCHTIEF ViCon GmbH} \\
Essen, Germany \\
alexandros.chirtoglou@hochtief.de}
\and
\IEEEauthorblockN{5\textsuperscript{rd} Christoph Schimanski}
\IEEEauthorblockA{\textit{HOCHTIEF ViCon GmbH} \\
Essen, Germany \\
christoph.schimanski@hochtief.de}
\and
\IEEEauthorblockN{6\textsuperscript{rd} Heja Goli}
\IEEEauthorblockA{\textit{HOCHTIEF ViCon GmbH} \\
Essen, Germany \\
heja.goli@hochtief.de}
}

\DeclareTextFontCommand{\emph}{\bfseries}

\begin{document}

\maketitle

\copyrightnotice

\thispagestyle{plain}
\pagestyle{plain}

\begin{abstract}
    \textit{Context and motivation:} 
    Classification aids software development activities by organizing requirements in classes for easier access and retrieval. The majority of requirements classification research has, so far, focused on binary or multi-class classification. \textit{Question/problem:} Multi-label classification with large taxonomies could aid requirements traceability but is prohibitively costly with supervised training. Hence, we investigate zero-short learning to evaluate the feasibility of multi-label requirements classification with large taxonomies. \textit{Principal ideas/results:} We associated, together with domain experts from the industry, 129 requirements with 769 labels from taxonomies ranging between 250 and 1183 classes. Then, we conducted a controlled experiment to study the impact of the type of classifier, the hierarchy, and the structural characteristics of taxonomies on the classification performance. The results show that: (1) The sentence-based classifier had a significantly higher recall compared to the word-based classifier; however, the precision and F1-score did not improve significantly. (2) The hierarchical classification strategy did not always improve the performance of requirements classification. (3) The total and leaf nodes of the taxonomies have a strong negative correlation with the recall of the hierarchical sentence-based classifier. \textit{Contribution:} We investigate the problem of multi-label requirements classification with large taxonomies, illustrate a systematic process to create a ground truth involving industry participants, and provide an analysis of different classification pipelines using zero-shot learning. 
\end{abstract}

\begin{IEEEkeywords}
requirements classification, domain-specific taxonomy, large-scale, multi-label
\end{IEEEkeywords}

\section{Introduction}\label{sec:intro}
In the infrastructure and construction domain, taxonomies are used as a means to structure and link information. Digital twins~\cite{tao2018digital} have the purpose of mirroring real-world objects, allowing engineers to plan, simulate, handshake implementation proposals with developers, and eventually hand over digital representations of constructions to their owners for operation and maintenance~\cite{jones2020characterising}. Taxonomies have in this context many use cases: streamlining terminology, structuring information, and providing unique identifiers for digital objects that have real-world counterparts.

\begin{figure*}[tb]
    \centering
    \includegraphics[width=\textwidth]{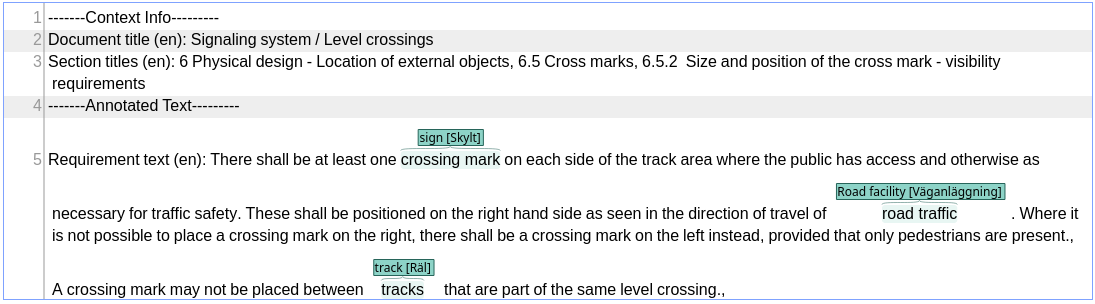}
    \caption{Example of a multi-label requirement classification}
    \label{fig:classification example}
\end{figure*}

With our research, we aim to add another use case for taxonomies in this context: enabling the tracing of requirements specifications to digital objects, which in turn enables requirements-based verification. Since manually assigning classes --- from taxonomies that can contain thousands of elements --- to thousands of requirements is not feasible in practice, we formulate this task as a (computer-supported) requirements classification problem with multiple labels.

Figure~\ref{fig:classification example} illustrates an example of multi-label requirement classification, which was performed by domain experts. The requirement has three classifications, \emph{sign}, \emph{road facility}, and \emph{track}, that come from a domain-specific taxonomy. These classifications are made based on the terms mentioned in the requirement text: \emph{crossing mark}, \emph{road traffic}, and \emph{tracks}. The document and section titles are examples of context information that help in understanding the context and classifying the requirement correctly. The classification of the terms alone can be seen as a multi-class problem. However, we are interested in classifying the requirements text as a whole, which can have multiple labels and therefore is a multi-label problem. 

Requirements classification is the second most researched NLP4RE task after defect detection~\cite{zhao2021natural}. Classification aids several requirements engineering (RE) activities, such as identifying non-functional requirements (NFR)~\cite{binkhonain2019review}, aiding regulatory compliance of software through ambiguity identification~\cite{massey_identifying_2014}, checking for requirements consistency and completeness~\cite{ott_automatic_2013} and traceability between requirements and software artifacts~\cite{mills_tracing_2019}.

Many studies focus on supervised binary classification (e.g., identification of NFR~\cite{kurtanovic_automatically_2017}) and multi-class classification (e.g., NFR classification~\cite{kurtanovic_automatically_2017}). 
Supervised machine learning requires a large amount of labeled data for model training which can be challenging to collect, in particular if labels need to be assigned manually. Furthermore, the transfer of trained classifiers to other domains is usually accompanied by a loss in performance~\cite{ott_automatic_2013}, making supervised models both expensive to create and less amenable for reuse. 

Supervised machine learning is therefore not a good solution fit for training a classifier on large taxonomies. They exhibit a large output space (the set of all possible classes in the taxonomy) and balanced training data would require hundreds or thousands of labeled examples for each class.

An alternative to supervised learning is zero-shot learning~\cite{larochelle_zero-data_2008}, which leverages the usage of pre-trained models to predict unseen classes~\cite{xian_zero-shot_2017,wang_survey_2019}. Zero-shot learning does not require any labeled data set for training, and the transfer of a classifier to a new domain does not require retraining~\cite{rezaei_zero-shot_2020}. However, to evaluate the
performance of a zero-shot learner, labeled data is still required. 

In previous work~\cite{unterkalmsteiner_early_2020}, we proposed to use a word2vec model to classify requirements using a domain-specific taxonomy. The performance of the implemented zero-shot classifier was not rigorously evaluated due to the lack of a ground truth, i.e., a set of correctly classified requirements that act as the gold standard for evaluation.

In this paper, we illustrate a process to construct a ground truth, involving industry practitioners to achieve high fidelity in the assigned labels. Furthermore, we report the results of an experiment that studies the effect of three variables on the performance of a multi-label requirements classifier. This paper's contributions are: (1) the illustration of a rigorous process for building a ground truth involving domain experts, (2) the development of a ground truth containing 129 requirements and 769 labels, (3) the analysis of the effect of three variables on the performance of the classifier, and (4) the correlation between output space characteristics and the performance of the classifier.

The remainder of the paper is structured as follows. Section~\ref{sec:background} contains the background information and the related work to our study. Section~\ref{sec:classification task} describes the requirements classification task and the classifiers we used in this experiment. We present the protocol of the experiment in Section~\ref{sec:experiment}. We answer the research questions in Section~\ref{sec:results}. In Section~\ref{sec:discussion}, we discuss the experiment results in relation to other studies. The conclusion of this paper is presented in Section~\ref{sec:conclusion}.

\section{Background and related work}\label{sec:background}
We briefly outline key concepts pertaining to our approach and review related work.

\subsection{Background}
\emph{Text classification} problems can be divided into binary, multi-class, and multi-label classification~\cite{sokolova_systematic_2009}. In binary classification, the output space consists of a positive and a negative class, e.g., the classification of requirements as functional or non-functional~\cite{kurtanovic_automatically_2017}. In multi-class classification, the output space contains more than two nominal classes, and an entity can fall into only one of these classes. For example, the classification of NFRs into one of the classes operability, performance, usability, or security~\cite{baker_automatic_2019} is a multi-class classification. Multi-label classification is similar to the multi-class, except an artifact can belong to multiple classes. For example, the classification of a GitHub repository into one or more topics~\cite{izadi_topic_2021} is a multi-label classification.

Another dimension to distinguish text classification problems is the structure of the output space. In hierarchical classification, the output space could be arranged in a tree or Direct Acyclic Graph~\cite{silla_survey_2011}. A distinct feature of hierarchical classification is the consideration of the relation between the different classes when deciding on the correct classification of an artifact~\cite{silla_survey_2011}. The classification is considered flat when information is classified without exploiting the relational information between the classes. A taxonomy is an example of an output space that consists of a controlled vocabulary, with a hierarchical structure~\cite{garshol_metadata_2004}. We explain taxonomies in more detail in Section~\ref{sec:class taxonomy}.

\emph{Zero-Shot learning} is a machine learning approach that uses pre-trained models to predict unseen classes~\cite{larochelle_zero-data_2008}. In text classification, zero-shot learning models are trained to learn the meaning of words on large corpora, e.g., generic ones like Wikipedia~\cite{qiao_less_2016} or domain-specific ones like StackOverflow~\cite{tabassum_code_2020}. These models use various NLP techniques (e.g., part-of-speech tagging and semantic embedding) to classify the text on yet unseen classes. Usually, the text is analyzed based on its semantic embedding, a vector representation of text in high dimensional vector space~\cite{jurafsky_speech_2009}. The vector can be sparse, generated based on the occurrence of a term in documents, term frequency-inverse document frequency (TF-IDF), or a dense vector, generated using neural network models (e.g., word2vec). Zero-shot learning is a potential alternative to supervised machine learning when a labeled dataset is difficult to obtain for model training, e.g., due to a large number of classes or the high cost of data labeling.

\subsection{Related work}\label{sec:related work}

Several researchers focused on binary classification in RE~\cite{casamayor_identification_2010,winkler_automatic_2016,kurtanovic_automatically_2017,alhoshan_zero-shot_2023,bashir2023requirement}, e.g., distinguishing requirements from non-requirements~\cite{winkler_automatic_2016,bashir2023requirement}. Interestingly, a classifier based on the BERT language model achieved F1-scores of 82\% on the positive class  (requirement) and 87\% on the negative class (non-requirement)~\cite{bashir2023requirement}, outperforming a supervised learning approach suggested by 
Winkler and Vogelsang~\cite{winkler_automatic_2016}. They achieved F1-scores of 80\% (positive) and 82\% (negative).

NFR identification and classification is one of the most studied classification problems in RE. It has been formulated as a binary classification (FR vs. NFR identification~\cite{casamayor_identification_2010, kurtanovic_automatically_2017, alhoshan_zero-shot_2023}), multi-class classification (identification of NFR types~\cite{kici_bert-based_2021}), and multi-label classification (multiple NFR types per information unit~\cite{ott_automatic_2013, sabir_multi-label_2020}) problem. 

Using supervised learning approaches, Kurtanovic et al.~\cite{kurtanovic_automatically_2017} achieved an F1-score of $\approx 92\%$ on the FR vs. NFR identification task, and between 51-82\% F1-score on the NFR classification task. Casamayor et al.~\cite{casamayor_identification_2010} could reduce the required training data while achieving a relatively good F1-score on the FR vs. NFR task. Nevertheless, the F1-score was lower on the NFR classification task, using a semi-supervised learning approach. Alhoshan et al.~\cite{alhoshan_zero-shot_2023} compared generic and domain-specific language models based on BERT (without fine-tuning) with supervised learning approaches on the FR vs. NFR task. Generic language models performed better than domain-specific language models and were comparable to supervised learning. However, overall, the supervised learning approach had the highest F1-score.

Kici et al.~\cite{kici_bert-based_2021} studied multi-class requirements classification on multiple dimensions: type (e.g., task, defect, or maintenance), severity, and priority. They compared transfer learning approaches through fine-tuning BERT models with other deep learning methods. They achieved an F1-score of up to 80\% when classifying requirements based on type, but lower based on severity and priority. Our research differs from that of Kici et al. since we classify requirements based on the topics (domain concepts) that the requirements cover. In addition, their approach requires labeled data to fine-tune BERT; however, we advocate for zero-shot learning, where no labeled data is used to train or fine-tune a model.

Ott~\cite{ott_automatic_2013} classified automobile requirements into one or more topics from a list of topics (multi-label), which are predetermined by a requirements engineer. The author uses two supervised learning approaches, multinominal naive Bayes and support vector machines, to identify relevant topics in a requirement. Even though he could achieve up to 83\% recall and 66\% precision on the training data, the transfer to another dataset led to a lower recall (40\%) and precision (50\%). He concluded that more training data is required for the classifier to achieve high performance on a dataset from a different project. Ott's aim was similar to ours in terms of classifying requirements using topics. It differs from ours, as we use a zero-shot learning approach without training the model on requirements from the same project. Sabir et al.~\cite{sabir_multi-label_2020} investigated the miss-classification of NFR requirements caused by automated techniques. They propose using deep learning approaches and assigning multiple labels to one requirement (multi-label) from a set of five classes, which is significantly smaller than the large taxonomies we are using.

Based on the literature discussed in this section, there is a lack of studies that 1) compare zero-shot learning approaches on a multi-label requirements classification task with large taxonomies, i.e. a large output space, and 2) investigate the effect of the output space characteristics on the classifier's performance. This study aims to fill this gap.

\section{Multi-label requirements classification with large taxonomies}
\label{sec:classification task}
In this section, we introduce the two classifiers that we evaluated. Furthermore, we explain the characteristics of taxonomies that we investigate in the experiment.

\subsection{Word-based classifier (word2vec)}
In a previous study~\cite{unterkalmsteiner_early_2020}, we proposed a word2vec-based classifier that classifies nouns in a requirement text; hence, we refer to it as the word-based classifier. 


\subsubsection{Preprocessing} The classifier uses a segmenter, tokenizer, stemmer and part-of-speech tagger to identify and extract nouns. Since we are interested in real-world entities that can be found in taxonomies, we focus our analysis only on nouns. We defined two predictors that estimate the degree of association between the nouns identified in the requirement and the nouns identified in the taxonomy's classes. The overall score for ranking the association between requirement and taxonomy nouns is the average score of the two predictors.
        
\subsubsection{Exact match predictor} If the exact same noun is found in the requirement and in the taxonomy, the score is:

\begin{equation}
    P_{exact} = \frac{1} {f_{noun}}
\end{equation}

where $f_{noun}$ refers to the number of taxonomy classes in which the noun appears. The score is lower the more frequent a noun is, i.e., the less distinguishing power between classes it has.
    
\subsubsection{Semantic similarity predictor} Instead of using Wikipedia as a general corpus, we trained a domain-specific word2vec model with data collected by a web search, instrumented by the nouns identified in the used taxonomy. For each noun in the requirement, the word2vec model is used to find the top 10 similar words (proxies). Similarity is calculated between each proxy and the taxonomy's classes using the equation:

    \begin{equation}\label{eq:ss predictor}
        P_{similarity} = \frac{1} {f_{proxy} * cos(\theta_{noun-proxy})}
    \end{equation}

where $f_{proxy}$ is the number of classes in which the proxy appears, and $cos(\theta_{noun-proxy})$ is the cosine similarity between the proxy and noun found in the requirement.

\subsection{Sentence-based classifier (ESA)}

The sentence-based classifier analyses the complete input text, not only the nouns. The main motive behind using a sentence-based classifier is its ability to distinguish between more than one possible word meaning based on the context. 

Gabrilovich et al.~\cite{gabrilovich_computing_2007} presented Wikipedia-based Explicit Semantic Analysis (ESA), which measures semantic relatedness between sentences of paragraphs, hence, the name sentence-based classifier. The main motive behind using a sentence-based classifier is its ability to distinguish between more than one possible word meaning based on the context~\cite{liu_topical_2015}. Song et al.~\cite{song2014dataless} built a dataless (zero-shot) classifier based on ESA to classify newswire stories and messages into a taxonomy of 26 topics with a depth level of two. This aligns with our problem as we classify requirements into domain concepts without labeled data. We adopted and modified their approach to classify natural language requirements based on the specified concepts. We explain the classification process next.

\subsubsection{Preprocessing} Stop-word removal is performed on the requirement's text and the taxonomy's nodes' description. Then, the nodes' descriptions are aggregated bottom-up where the child nodes' descriptions are brought up to the parent node. Furthermore, the requirement text with the document and section titles are aggregated. 

\subsubsection{Explicit semantic analysis (ESA)} A vector of concepts is built for a text fragment by identifying the relevant concepts from Wikipedia articles. The concepts in the vector are weighted using the TF-IDF score of the text~\cite{chang_importance_2008,song2014dataless}. This process is conducted on both the requirement text and the nodes in the taxonomy.

\subsubsection{Relatedness calculation} 

The semantic relatedness is measured between the requirement's vector and each of the taxonomy nodes' vector using cosine measure as follows:

\begin{equation}
    score = cos(\phi_{x}(l_{i}), \phi_{x}(r))
\end{equation}

where $\phi_{x}$ is the ESA representation of a text, $r$ is the aggregated requirement text, and $(l_{i})$ is node $i$ in the taxonomy. Then, the similarity score is normalized in the range $0\ldots 1$.

\subsubsection{Global classifier}
The classes of the taxonomy are sorted based on the relatedness score, and the top $k$ labels are selected.

\subsubsection{Adaptations}
The algorithm that we adopted~\cite{chang_importance_2008,song2014dataless} uses a local classifier per level, which classifies text in a given depth level of the taxonomy~\cite{silla_survey_2011}. We replaced the classifier with a global classifier, which produces $k$ labels regardless of the depth level. This is due to our interest in classifying requirements with the most relevant class at the deepest level, which represents the most concrete class.

\subsection{Classification with taxonomies}\label{sec:class taxonomy}

A taxonomy is a controlled vocabulary arranged in a hierarchy or an inverse tree~\cite{garshol_metadata_2004}. It could contain one or more dimensions, and each dimension can be used to classify artifacts from a different angle. The quality of a taxonomy can be characterized by various attributes~\cite{unterkalmsteiner2023compendium}. We study the effect of five structural characteristics on the requirements classification task. We chose these characteristics as they can be used to describe the size and shape of the output space, whose scale is a key differentiator of our study compared to related work.

\begin{enumerate}
    \item Description length: the average number of characters per class description, including name, description, and synonyms of the class.
    \item Depth: the maximum level of depth of the taxonomy.
    \item Categories: the number of intermediate classes with at least one child class.
    \item Leaf nodes: the number of classes on the lowest levels with no child classes.
    \item Total nodes: the total number of classes in the taxonomy equals the sum of categories and leaf nodes.
\end{enumerate}

In principle, there are no restrictions on the taxonomies that can be used for requirements classification. For practical purposes, we use in our experiment (Section~\ref{sec:experiment}) domain-specific taxonomies that are in use at the companies involved in this research. 

\section{Experiment Planning}\label{sec:experiment}
We report the design and results of our experiment following the guidelines for reporting experiments in software engineering~\cite{jedlitschka_reporting_2008}.

\subsection{GQM}\label{sec:GQM}
We start by constructing a GQM (Goal-Question-Metric) matrix. The aim of the experiment is to understand the impact of certain design decisions of the classification pipeline and differences of the taxonomy characteristics on the classification performance when using artifacts that originate from practice. We break the aim down into three main goals.

\begin{enumerate}
     \item G1: Evaluate the impact of classifier type on classification performance. 
     \item G2: Evaluate the impact of exploiting a taxonomy's hierarchical structure on the classification performance.
     \item G3: Evaluate the impact of a taxonomy's characteristics on the classification performance.
\end{enumerate}

In alignment with our goals, we define the following research questions.

\begin{enumerate}
     \item RQ1: To what extent does the word-based and sentence-based classifier performance differ?
     \item RQ2: To what extent does the hierarchical classification strategy differ from the flat classification strategy in terms of performance?
     \item RQ3: To what extent do structural characteristics of the taxonomy impact classifier performance?
\end{enumerate}

Table~\ref{tab:metrics} presents the performance metrics that we use to answer the questions: 1) recall, which represents the sensitivity; 2) precision, which represents specificity; and 3) F1-score, the harmonic mean between the former two. These are common metrics used to measure the performance of a classifier~\cite{kowsari_text_2019}. We calculated these metrics using micro averages. We motivate the use of micro averages in Section~\ref{sec:analysis procedure}.

\begin{table}[ht]
    \centering
    \footnotesize
    \caption{The metrics used to measure the performance of the classifiers}    
    \begin{tabular}{ccc}
        \toprule
        \textbf{Id} & \textbf{Metric} & \textbf{Equation} \\
        \midrule
        1 & Recall      &   $\frac {\sum_{i=1}^{l} tp_{i}} {\sum{i=1}^{l} (tp_{i} + fn_{i})}$
        \\[10pt]
        2 & Precision   &   $\frac {\sum_{i=1}^{l} tp_{i}} {\sum{i=1}^{l} (tp_{i} + fp_{i})}$
        \\[10pt]
        3 & F1          &   $2 * \frac { Precision * Recall} { Precision + Recall}$
        \\[10pt] \bottomrule
    \end{tabular}
    \begin{tablenotes}
        \item[a] tp = true positive, tn = true negative, fp = false positive, fn = false negative
    \end{tablenotes}
    \label{tab:metrics}
\end{table}

\subsection{Materials}
\label{sec:materials}

The experiment materials consist of natural language requirements and the taxonomies used to classify these requirements.

\subsubsection{Requirements}
A total of 129 requirements were sampled by the second author. The sample consists of 92 regulatory requirements from publicly available documents\footnote{\url{https://puben.trafikverket.se/dpub/sok}} and 37 project-specific, non-public, requirements. The length of the sampled requirements text ranges between 45 and 856 characters. We used the requirement text together with context information (document and section titles) for classification.

\subsubsection{Taxonomies} 
We used two taxonomies (SB11 and CoClass), each containing three dimensions that describe orthogonal aspects of the problem domain (construction). Both taxonomies were developed by a consortium of Swedish construction companies and the Swedish Transport Administration.
We choose these taxonomies since their dimensions have varying structural characteristics as presented in Table~\ref{tab:cs_ch}, which helps us to answer Q3. In our experiment, we treat each dimension as a separate output space and, consequently, have six output spaces in total. 

\begin{table}[h]
    \caption{Output space characteristics}\label{tab:cs_ch}
    \centering
    \begin{threeparttable}
    \footnotesize
    \resizebox{1\columnwidth}{!}{
    \begin{tabular}{lllllll}
        \toprule
             \textbf{Output space} & \textbf{Desc. Length} & \textbf{Depth} & \textbf{Categories} & \textbf{Leaf Nodes} & \textbf{Total Nodes} \\
        \midrule
             $OS_{A}$ & 27 & 5 & 50 & 206 & 256 \\
             $OS_{B}$ & 28 & 6 & 299 & 884 & 1183 \\
             $OS_{G}$ & 96 & 3 & 199 & 665 & 864 \\
             $OS_{K}$ & 92 & 3 & 61 & 251 & 312  \\
             $OS_{L}$ & 40 & 1 & 0 & 635 & 635 \\
             $OS_{T}$ & 79 & 4 & 80 & 170 & 250 \\
        \bottomrule
        \end{tabular}
    }
    \begin{tablenotes}
        \item[a] Desc. length: mean characters count of node description
    \end{tablenotes}
    \end{threeparttable}
\end{table}

\subsection{Building the Ground Truth.}

\begin{figure*}[bth]
    \centering
    \includegraphics[width=1\textwidth]{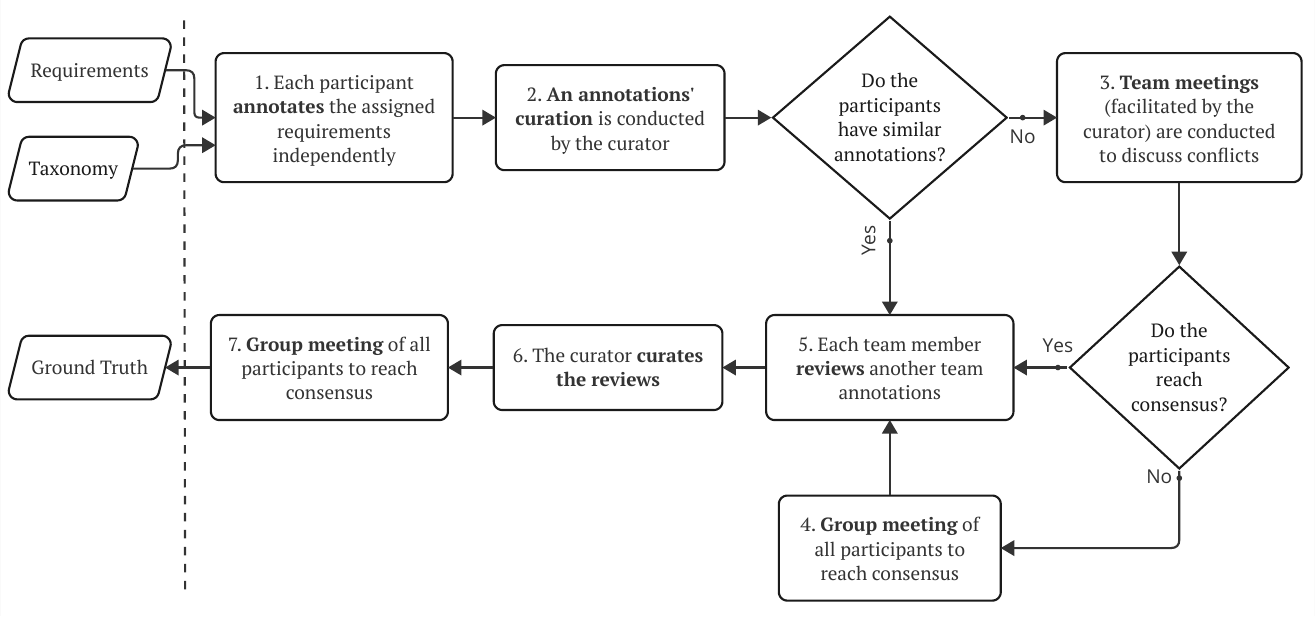}
    \caption{The process of building the ground truth (UML 2.0)}
    \label{fig:ground truth}
\end{figure*}

The measurement of the classifier's performance requires a ground truth, that is, a set of requirements correctly labeled using the classes from a taxonomy. Creating such a ground truth is challenging due to the large output space. Even domain experts struggle to navigate the taxonomies and select the ``correct'' classes. The question of whether a selected class is correct or not is a matter of agreement between the experts. Due to this difficulty, we designed a process that captures this search for agreement in a systematic, controlled manner (Figure~\ref{fig:ground truth}). 

\begin{table}[htb]
    \centering
    \footnotesize
    \caption{Participants in building the ground truth with years of experience in the domain and in research}
    \begingroup
    \setlength{\tabcolsep}{3pt} 
    \begin{tabular}{cllcc}
    \toprule
        \textbf{Team} &
        \textbf{Affiliation} & 
        \textbf{Role} & 
        \textbf{Domain} &
        \textbf{Research}\\
    \midrule
        A & University & Researcher & none & 13 \\
        A & Company & Civil engineer & $\prec$ 1 & $\prec$ 1 \\
        B & University & Researcher (curator) & none & 3 \\
        B & Company & BIM manager and consultant & 5 & 4 \\
        C & University & Researcher & none  & 14 \\
        C & Company & BIM and information manager & 7 & 1\\ \bottomrule
    \end{tabular}
    \endgroup
    \label{tab:participants}
\end{table}

Six participants, forming three teams of two, were involved in building the ground truth. The participants' team, roles, and years of experience in the (construction) domain and research are listed in Table~\ref{tab:participants}. The participants used INCEpTION~\cite{tubiblio106270}, a platform used for annotation tasks, to annotate the sampled requirements with the classes originating from the six output spaces presented in Section~\ref{sec:materials}. In total, we created 769 labels for the 129 requirements. We explain next the steps of the process shown in Figure~\ref{fig:ground truth}.

\subsubsection{Annotation (all, individually)} Each participant annotated a set of assigned requirements. An annotation is made by finding the appropriate node(s) (1 or more) from each output space to a token, a word (usually nouns), a phrase, or a sentence. In case no appropriate node was found for a token, a node called "No Available Code" was used. 
    
\subsubsection{Annotation Curation (curator)} The curator examined the requirements' annotations created by each team member and judged if the annotations were equivalent within the same team. Annotations are considered equivalent if the same node from the output space is used to annotate identical text parts. Any errors caused by the misuse of the tool were fixed by the curator. If a requirement had similar annotations by both team members, the requirement was moved to step 5: the review process. Otherwise, the requirement was scheduled for discussion in step 3: team meetings.
    
\subsubsection{Team meetings (team + curator)} Team meetings were conducted and facilitated by the curator. The curator presented the requirements and the annotations made by each team member and asked them to rationalize their annotations. If the team members agreed, the requirement was moved to step 5: the review process. Otherwise, the requirement was set for discussion in step 4: group meeting.
    
\subsubsection{Group meeting (all)} A group meeting was conducted to discuss and reach a consensus about the requirements with misaligned annotations from step 4: team meetings. The curator presented the requirement, and the annotators were asked to rationalize their annotations to make a consensus decision.

\subsubsection{Review (all, individually)} The annotated requirements by each team were assigned by the curator in a round-robin fashion to another team for review. Each team member was asked to review (individually) the requirements that were annotated by another team.

\subsubsection{Review Curation (curator)} The curator inspected the reviewers' comments and prepared a list of requirements that needed to be discussed in the group meeting. 

\subsubsection{Group meeting (all)} A final group meeting was conducted with all participants. In this meeting, the curator led the meeting by introducing the requirements, one by one, and asking the annotators and reviewers for that requirement to discuss their points of view. The remaining group members could also participate in the discussion.

We performed this process in four rounds, annotating around 30 requirements in each. We chose to split up the annotation into rounds to be able to modify the process if necessary (we did not) and to gauge whether we improved in our agreement over time (we did). The inter-rater reliability (IRR) between the reviewers and annotators is shown in Table~\ref{tab:irr}. In the first round, the average IRR was 64\%. In the subsequent rounds, the average IRR improved and reached 89\% in the final round. This is an indication that the classification task is not trivial and expertise is required to ensure a high-quality classification. The increase of the IRR in the subsequent rounds can be attributed to the learning effect after the participants have conducted the task four times.

\begin{table}[ht]
    \caption{Ground truth inter-rater reliability}\label{tab:irr}
    \centering
    \begin{threeparttable}
    \footnotesize
    \begingroup
    \setlength{\tabcolsep}{3pt} 
    \begin{tabular}{llllll}
    \toprule
         \textbf{Review round} & \textbf{Reviewers} & \textbf{Annotators} & \textbf{Spans} & \textbf{IRR} & \textbf{Average} \\
        \midrule
        \multirow{3}{*}{1} &  A &	C & 25 & 52.00\%  & \multirow{3}{*}{64.60\%} \\
         &   B & A & 32 & 62.50\% & \\
         &	C & B & 25 & 80.00\% & \\  
         \midrule
        \multirow{3}{*}{2} &  A &	B & 20 & 90.00\% & \multirow{3}{*}{83.66\%} \\
        &	B & C & 27 & 77.78\% & \\
        &	C & A & 38 & 84.21\% & \\
        \midrule
        \multirow{3}{*}{3}  &  C &	B &	37 & 83.78\% & \multirow{3}{*}{83.00\%} \\
        &	A & C &	29 & 86.21\% & \\
        &	B & A & 31 & 77.42\% & \\
        \midrule
        \multirow{3}{*}{4} & C &	A & 33 & 87.88\% & \multirow{3}{*}{89.00\%} \\
        &	A &B& 26 & 96.15\% & \\
        &	B &	C & 22 & 86.36\% & \\
    \bottomrule
    \end{tabular}
    \endgroup
    \begin{tablenotes}
        \item IRR: Inter-rater reliability calculating using joint probability of agreement between the reviewers and the annotators.
    \end{tablenotes}
    \end{threeparttable}
\end{table}

Table~\ref{tab:ground truth} depicts the number of requirements annotated per output space and the number of labels that we used to classify these requirements. We refer to each set of requirements that we labeled with an output space as a sample. In the end, we created 769 labels for the requirements, which took around 446 person-hours (including sampling and meetings).

\begin{table}[htb]
    \caption{Ground truth characteristics}\label{tab:ground truth}
    \centering
    \begin{threeparttable}
    \footnotesize
    \begin{tabular}{llll}
    \toprule
        \textbf{Sample Id} & \textbf{Requirements} & \textbf{Output Space} & \textbf{Labels} \\
    \midrule
        S1 & 24 & $OS_A$ & 31  \\
        S2 & 123 & $OS_B$ & 254 \\
        S3 & 98 & $OS_G$ & 205 \\
        S4 & 59 & $OS_K$ & 74  \\
        S5 & 56 & $OS_L$ & 88  \\
        S6 & 81 & $OS_T$ & 117  \\
    \midrule
        Total & & & 769 \\
    \bottomrule
    \end{tabular}
    \end{threeparttable}
\end{table}

\subsection{Tasks}\label{sec:tasks}
Both classifiers assign a score to each label and return an ordered list of labels. In practical use, such a classifier would return $k$ labels and an engineer would choose from this set the most appropriate labels for the classified requirement. If $k$ is too small, correct labels that are ranked lower would be omitted from the result. If $k$ is too large, an engineer would have to review a large amount of irrelevant labels. This is the typical trade-off between optimizing precision and recall. In this study, we set $k=15$. With an output space of 250 - 1183 labels, this would mean for an engineer a reduction of 94\% and 99\% of the labels, respectively. In practice, it would certainly make sense to experiment with other values of $k$, in particular increasing it to achieve the highest recall under a tolerable workload.

Due to the design of the classifiers, setting $k$ to a fixed value makes their performance incomparable. The word-based classifier produces $j$ labels \emph{per noun}, whereas the sentence-based classifier produces $j$ labels \emph{per requirement}. If we set $k=15$, and identify three nouns in a requirement, we would evaluate the first 45 labels (for each noun, 15 labels) for the word-based classifier and only 15 for the sentence-based classifier. Therefore, to make the classifiers comparable, we normalize the output of the word-based classifier by limiting the labels per noun to $l_{noun} = \frac{k}{m}$, where $m$ is the number of nouns in the requirement. 
For example, considering $k = 15$ and requirement $R$ has four nouns $m = 4$, so $l_{noun} = \frac{15}{4} = 3.75$. We take the top 4 labels per noun and discard the label with the lowest score to end up with 15 labels, i.e. the same number we get with the sentence-based classifier.

\subsection{Design, Hypotheses, and Variables}

\renewcommand{\theenumi}{\Alph{enumi}}

The design of the experiments is 2x2x6 (24), i.e., two factors (A, B) with two treatments each and one factor (C) with six treatments~\cite{wohlin2012experimentation,montgomery2017design}:
\begin{enumerate}
    \item The classifier type, with treatments word-based and sentence-based. 
    \item Hierarchical strategy, with treatments flat and hierarchical. The flat strategy does not consider the structure of the output space when processing it. The hierarchical strategy follows a bottom-up approach, i.e. the description of each node (i.e., class) in the output space is aggregated to the parent until the descriptions of all nodes are in the root.
    \item Output space with six treatments, $OS_{A}$, $OS_{B}$, $OS_{G}$, $OS_{K}$, $OS_{L}$ and $OS_{T}$. The description of these output spaces is in Section~\ref{sec:materials}.
\end{enumerate}

The dependent variable is the classifier performance, which is measured based on the metrics presented in Table~\ref{tab:metrics}. We generate 42 hypotheses following the templates in Table~\ref{tab:hypotheses}, where $M$ is the metric value (recall, precision, F1-score), $CH$ represents the output space characteristics (description length, depth, categories, leaf nodes, and total nodes), and $corr$ is the correlation factor between the metrics and characteristics.

\begin{table*}[bth]
    \centering
    \caption{Hypotheses Formulas}
    \resizebox{1\textwidth}{!}{
    \begin{tabular}{cccc}
        \toprule
        factor & factor A & factor B & factor C  \\
        \midrule
        hypotheses
        &
        $H_{0A}: M(Word) = M(Sentence)$
        &
        $H_{0B}: M(Flat) = M(Hierarchy)$
        &
        $H_{0C}: corr(M, CH) <= 0$
        \\
        &
        $H_{1A}: M(Word) \neq M(Sentence)$
        & 
        $H_{1B}: M(Flat) \neq M(Hierarchy)$
        &
        $H_{1C}: corr(M, CH) > 0$
        \\
        \midrule
        count &
        2 * 3 metrics = 6 &
        2 * 3 metrics = 6 &
        2 * 3 metrics * 5 chars = 30
        \\
        \bottomrule
    \end{tabular}
    }
    \label{tab:hypotheses}
\end{table*}

\subsection{Analysis Procedure}\label{sec:analysis procedure}

We test the hypotheses using non-parametric tests due to the small sample size where normal distribution can not be assumed. We assigned the same requirements sample to all combinations of factor A x factor B, referred to as paired-comparison~\cite{wohlin2012experimentation}. Thus, we use the Wilcoxon test to analyze the effect of factor A (answering RQ1) and factor B (answering RQ2) on the classifier's performance. 
As for factor C, we calculate Spearman's correlation coefficient~\cite{myers_spearman_2006} between the metrics and the output space characteristics (answering RQ3).

A classifier's performance metrics (precision, recall, F1-score) can be calculated using micro averages (per document) and macro averages (per class)~\cite{yang_evaluation_1999,sun_hierarchical_2001}. By analyzing the collected ground truth, we observed an imbalance in the number of samples per class. In such a situation, micro or weighted average should be used~\cite{kici_bert-based_2021}. The motivation is as follows. The macro average is calculated across the classes in the output space. Thus, all the classes should be present in the dataset for the metric to be representative of the classifier's performance across the whole output space. In our case, not all classes are present in the dataset due to the large output space. Consequently, we calculate and report the micro average as it represents a per-document average.

\section{Results}
\label{sec:results}

Table~\ref{tab:classifiers_comparison} depicts the results of the classifier performance. We answer the research questions by following the analysis procedure described in Section~\ref{sec:analysis procedure}. 

\begin{table}[tbh]
    \caption{Classification performance of all factor combinations} 
    \footnotesize\label{tab:classifiers_comparison}
    \begin{threeparttable}
    \resizebox{1\columnwidth}{!}{
    \begingroup
    \setlength{\tabcolsep}{10pt} 
    \begin{tabular}{llllll}
    \toprule
        Classifier type & Hierarchy & OS & Recall & Precision & F1-score \\ \midrule
        word-based & flat & A & 0.37 & \underline{\textbf{0.07}} & \underline{\textbf{0.11}} \\ 
        sentence-based & flat & A & 0.4 & 0.03 & 0.06 \\ 
        word-based & hierarchical & A & 0.13 & 0.03 & 0.05 \\ 
        sentence-based & hierarchical & A & \textbf{0.53} & 0.04 & 0.08 \\ 
        \midrule
        word-based & flat & B & 0.16 & 0.03 & 0.05 \\ 
        sentence-based & flat & B & 0.22 & 0.03 & 0.05 \\ 
        word-based & hierarchical & B & 0.11 & 0.03 & 0.04 \\ 
        sentence-based & hierarchical & B & \textbf{0.37} & \textbf{0.05 }& \textbf{0.08} \\ 
        \midrule
        word-based & flat & G & 0.26 & 0.04 & 0.07 \\ 
        sentence-based & flat & G & \textbf{0.39} & \textbf{0.05} & \textbf{0.08} \\ 
        word-based & hierarchical & G & 0.08 & 0.01 & 0.03 \\ 
        sentence-based & hierarchical & G & 0.37 & 0.04 & \textbf{0.08} \\ 
        \midrule
        word-based & flat & K & 0.1 & 0.01 & 0.03 \\ 
        sentence-based & flat & K & \textbf{0.64} & \textbf{0.05} & \textbf{0.09} \\ 
        word-based & hierarchical & K & 0.1 & 0.02 & 0.03 \\ 
        sentence-based & hierarchical & K & 0.57 & \textbf{0.05} & 0.08 \\ 
        \midrule
        word-based & flat & L & 0.21 & 0.03 & 0.06 \\ 
        sentence-based & flat & L & \textbf{0.53} & \textbf{0.05} & \textbf{0.09} \\ 
        word-based & hierarchical & L & 0.21 & 0.03 & 0.06 \\ 
        sentence-based & hierarchical & L & \textbf{0.53} & \textbf{0.05} & \textbf{0.09} \\ 
        \midrule
        word-based & flat & T & 0.37 & \underline{\textbf{0.07}} & \underline{\textbf{0.11}} \\ 
        sentence-based & flat & T & 0.49 & 0.04 & 0.08 \\ 
        word-based & hierarchical & T & 0.03 & 0.01 & 0.01 \\ 
        sentence-based & hierarchical & T & \underline{\textbf{0.66}} & 0.06 & \underline{\textbf{0.11}} \\ \bottomrule
    \end{tabular}
    \endgroup
    }
    \begin{tablenotes}
        \item OS: output space. Bold scores: the highest scores on each output \\space. Underlined scores: the highest scores across all experiments.
    \end{tablenotes}
    \end{threeparttable}    
\end{table}

\subsection{RQ1: Word-based vs. Sentence-based}
The results of the Wilcoxon test on the classifier type factor data are:

\textit{Recall.} $\{T^-=78, T^+=0\}$, at $\alpha$ = 0.05, the results are significant as $W = min(T^-,T^+) = 0 < 13$, the critical value at $N = 12$. Consequently, we reject the null hypothesis $H_{01A}$ for recall.

\textit{Precision.} $\{T^-=49.5, T^+=16.5\}$, at $\alpha$ = 0.05, the results are not significant as $W = min(T-,T+) = 16.5 > 10$, the critical value at $N = 11$. Consequently, we can not reject the null hypotheses $H_{01A}$ for precision.

\textit{F1-score.} $\{T-=54.5, T+=11.5\}$, at $\alpha$ = 0.05, the results are not significant as $W = min(T-,T+) =  11.5 > 10$, the critical value at $N = 11$. Consequently, we can not reject the null hypotheses $H_{01A}$ for F1-score.

Based on the statistical tests and the results in Table~\ref{tab:classifiers_comparison}, the recall of the sentence-based classifier was significantly higher than the recall of the word-based classifier. However, the precision and F1 score did not differ significantly.

\subsection{RQ2: Hierarchical vs. Flat}

The results of the Wilcoxon test on the hierarchy factor data are:

\textit{Recall.} $\{T^-=15, T^+=30\}$, at $\alpha$ = 0.05, the results are insignificant as $W = min(T^-,T^+) = 15 > 5$, the critical value at $N = 9$. Consequently, we can not reject the null hypotheses $H_{01B}$ for recall.

\textit{Precision.} $\{T^-=13, T^+=23\}$, at $\alpha$ = 0.05, the results are not significant as $W = min(T-,T+) = 13 > 3$, the critical value at $N = 8$. Consequently, we can not reject the null hypotheses $H_{01B}$ for precision.

\textit{F1-score.} $\{T-=12, T+=24\}$, at $\alpha$ = 0.05, the results are not significant as $W = min(T-,T+) =  12 > 3$, the critical value at $N = 8$. Consequently, we can not reject the null hypotheses $H_{01A}$ for F1-score.

The hierarchical classifier did not perform consistently better than the flat classifier across all the experiments that we conducted.

\subsection{RQ3: The Correlation between the output space characteristics and classification performance}

\begin{figure*}[bth]
    \centering    \includegraphics[width=0.9\textwidth]{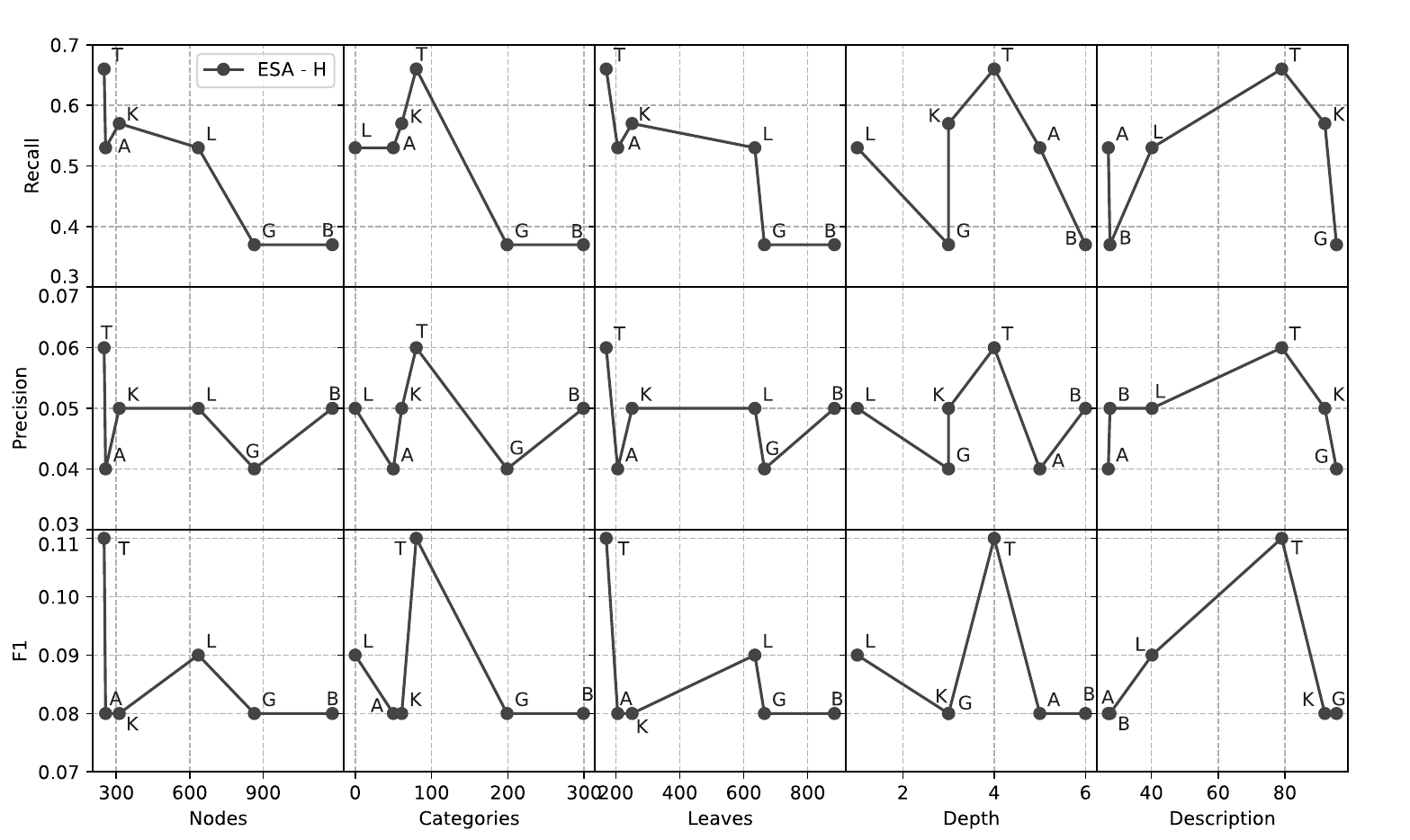}
    \caption{Correlation between output space characteristics and the sentence-based hierarchical classifier performance. Each point corresponds to one output space.}
    \label{fig:correlation}
\end{figure*}

Figure~\ref{fig:correlation} depicts the correlation between the output space characteristics and the sentence-based hierarchical classifier performance. The x-axis of each sub-figure represents a characteristic of the output space, while the y-axis represents a performance metric. One data point in a line plot corresponds to one output space. We did not calculate Sparman's rank correlation coefficient on the other factors combinations as the flat strategy was not affected by characteristics of the output space, and the word-based classifier has shown degraded performance when implementing the hierarchical strategy. 

\textit{Total Nodes and Leaf Nodes.}
According to Spearman's test, recall correlates with the total and leaf nodes. We reject $H_{0c}$ for (recall, leaf nodes) and (recall, total nodes). Looking at Figure~\ref{fig:correlation}, recall decreased as the number of total or leaf nodes increased except for one case where the number of total nodes increased from 256 ($OS_{A}$) to 312 ($OS_{K}$), and leaf nodes increased from 206 ($OS_{A}$) to 251 ($OS_{K}$), at that point the recall increased slightly by 0.04. The Spearman's test shows no correlation between precision or F1-score, and total or leaf nodes. Also, the differences in precision and F1 are minimal (0.01 - 0.03). 

\textit{Categories, Depth, and Description.}
There is no correlation between the number of [categories, depth, description] and [recall, precision, or F1-score], according to Spearman's test. 

\section{Discussion}\label{sec:discussion}


Compared to the results by Song et al.~\cite{song2014dataless}, which we used as inspiration for our sentence-based hierarchical classifier implementation, we achieved an overall lower performance, especially in terms of precision and F1-score. This can be attributed to the differences in the experimental materials. First, Song et al. used a small set of 26 topics to classify the documents. We used output spaces containing between 250 and 1183 nodes which is between one and two orders of magnitude larger than Song et al.'s. Second, the documents that Song et al. classified contain more text (a couple of paragraphs) compared to the requirements' text (a couple of sentences) that we classified. Third, in their study, a document belongs to one topic (multi-class), while in ours a requirement has multiple topics (multi-label). These differences potentially explain the lower performance we observed in the experiments as the classification tasks are more challenging. 

The performance obtained by Ott~\cite{ott_automatic_2013} (0.79 recall, 0.16 precision) on a similar multi-label requirements classification task (141 labels, which is at the lower end w.r.t. to the output space we experimented with) is slightly higher than the best performance that we obtained (0.66 recall, 0.06 precision). Ott used a supervised learning classifier, multinomial naive Bayes combined with 4-gram indexing and topic generalization, while we used a zero-shot hierarchical classifier. The proximity of the classifiers' performance demonstrates the potential of zero-shot classifiers to perform as well as the supervised learning classifiers without the time-consuming labeling of requirements. 

The word-based classifier exhibited a lower performance than the sentence-based classifier. This confirms the hypothesis originating from the qualitative feedback we received in our previous field experiment with practitioners~\cite{unterkalmsteiner_early_2020}: extracting and considering nouns individually removes important information.

The best recall (0.66) of the hierarchical sentence-based classifier was obtained using the output space $OS_{T}$ (nodes = 250, categories = 80, leaf nodes = 170, depth = 4, and description length = 79). In comparison to the other output spaces, $OS_{T}$ has the least number of nodes, an average number of categories and depth, and a relatively long description per node. 
The worst recall (0.37) of the hierarchical sentence-based classifier was obtained using the output spaces $OS_{G}$ (nodes = 864, categories = 199, leaf nodes = 665, depth = 3, and description length = 96) and $OS_{B}$ (nodes = 1183, categories = 299, leaf nodes = 884, depth = 6, and description length = 29). These are the two largest output spaces compared to the others in terms of the number of nodes, categories, and leaf nodes.
The highest performance value was obtained at a near-average depth level of 4 and description length of 79 (Figure~\ref{fig:correlation}). In contrast, the performance was lower at the remaining depth levels and description lengths. We attribute this to the hierarchical classification strategy where the description of the nodes is aggregated bottom up. A higher depth level and description length mean a longer text per class to analyze, which increases the noise and consequently makes the classification difficult. While at lower values of depth level and description length, we do not have enough description per class to classify correctly, an essential component of zero-shot learning to predict unseen classes~\cite{wang_survey_2019}.

\subsection{Implications on practice}
Efficiently classifying requirements using domain-specific taxonomies helps to manage (e.g., analyze and trace) requirements w.r.t. other artifacts. Such classification can be used for multiple purposes, e.g., traceability of requirements across multiple artifacts, assignment of requirements to teams based on concepts, or support creating evidence for safety-critical systems, i.e., that they are developed in accordance with the required regulations and laws. 

Adopting a zero-shot learning approach for requirements classification is more cost-efficient than supervised learning approaches due to the effort associated with building a training set. Instead of building a large dataset (over 26.000 data points for 141 classes~\cite{ott_automatic_2013}) to train a supervised learning model, one could prepare a high-quality ground truth that is significantly smaller and can be used to evaluate a pre-trained model.
Ott has created approximately 200 data points per class for training. If we take the smallest taxonomy from our set with 250 classes, we would need 50.000 data points for training. Extrapolating from the effort data for creating our ground truth, labeling such a training set would require over 29.000 person/h (or 10 people, annotating for three years, eight hours per day).

\subsection{Threats to Validity}

We discuss four aspects of validity, namely construct, internal, external, and conclusion validity~\cite{jedlitschka_reporting_2008,wohlin2012experimentation}.

\textit{Construct validity.} A threat to the construct validity is the usage of per-document averages for recall, precision, and F1 to measure the performance of the classifier. The per-document average does not show the prediction performance in all the classes. Instead, a per-class average should be measured, which we could not measure as not all classes are present in our ground truth. 

\textit{Internal Validity.} One of the threats is the potential bias that could be introduced during the execution of the study. To mitigate this threat, we randomly sampled the data and published a replication package containing the source code and the dataset used in the experiments. However, 35 out of the 129 requirements can not be published due to their proprietary nature. 

\textit{External validity.} The classification of requirements using domain taxonomy is dependent on the domain. Thus, there is a threat to the generalization of results in the software domain. More validation in different domains is required. Another possible threat to external validity is the size of our requirements set. We can not claim that the 129 requirements that we sampled represent all the requirements in a large infrastructure project. 

\textit{Conclusion validity.} Our study's conclusion is dependent on the selection and application of statistical tests, which we selected following the recommendations by Wohlin et al.~\cite{wohlin2012experimentation}. We used the Wilcoxon test, a non-parametric test, which is more robust against violations of assumptions about the data (i.e., non-normal distribution) than a parametric test. 

\section{Conclusion}\label{sec:conclusion}

We have evaluated the impact of word-based and sentence-based classifiers, flat and hierarchical classification strategies, and six large taxonomies on the performance of multi-label requirements classification. To enable this comparison, we have built a ground truth using a rigorous process involving domain experts. Using a sentence-based classifier, combined with the hierarchical strategy, has significantly improved the classification performance compared to the word-based classifier that we developed earlier. However, the absolute performance of the classifier is not yet sufficient to be used in practice as the precision is still low. In practice, it would still require considerable effort by an engineer to choose the correct labels. The recall is likely not high enough to induce confidence in retrieving all relevant labels.

Multi-label requirements classification is difficult both for humans and machines, in particular with large taxonomies. Therefore, a trade-off needs to be made between the granularity of classifications and the percentage of the correctly classified artifacts, depending on the application of the classification. The results we achieved are encouraging and may lead to more studies in multi-label requirements classification as it potentially enables other activities, e.g., fine-grained traceability. In future work, we plan to focus on improving the precision and overall performance of the classifier by introducing more pre-processing steps and using other zero-shot learning classifiers, e.g., using BERT and other large language models.

\section{Acknowledgments}

The authors would like to thank Dr. Alessio Ferrari (CNR-ISTI, Pisa, Italy) for his valuable input on this study.

\section{Data Availability}

We provide our replication package including the source code and the dataset here~\cite{dataset}. We have published 94 labeled requirements from the publicly available data. The remaining 35 requirements are project-specific and are not published due to the protection of company data. However, the remaining requirements can be made available upon a reasonable request.

\bibliographystyle{splncs04}
\bibliography{main}

\end{document}